\documentclass[preprint,showpacs,preprintnumbers,amsmath,amssymb,letterpaper]{revtex4}

\usepackage{graphicx}% Include figure files
\usepackage{dcolumn}% Align table columns on decimal point
\usepackage{bm,bbm}% bold math
\usepackage[english]{babel}
\usepackage{amssymb,amsmath,amsthm,newlfont,amsfonts,amsxtra,color}
\usepackage[latin1]{inputenc}
\usepackage{pifont}

\DeclareMathOperator{\diag}{diag}

\DeclareMathOperator{\Ber}{Ber}

\def\dalpha{\dot{\alpha}}

\newcommand{\iim}{\mathrm{i}}
\newcommand{\Oh}{\mathrm{O}}

\theoremstyle{plain}

\newtheorem{prop-en}{Proposition}

\theoremstyle{definition}

\newtheorem{Preg}{$\bullet$: Pregunta}[section]

\begin{document}

\preprint{ITP-UH-13/04}

\title{${\mathcal{N}}=2$ Super-Born-Infeld from \\ 
Partially Broken  ${\mathcal{N}}=3$ 
Supersymmetry in $d=4$}

\author{A. De Castro}
 \altaffiliation[Also at ]{Instituto Venezolano de Investigaciones Científicas,
 Centro de Física, Apartado Postal 21827-A, Caracas, Venezuela}
 \email{castro@itp.uni-hannover.de}
\author{L. Quevedo}
 \email{quevedo@itp.uni-hannover.de}
\affiliation{Institut für Theoretische Physik\\
Universität Hannover\\
Appelstraße 2, 30167 Hannover, Germany}

\author{A. Restuccia}
\affiliation{Universidad Simón Bolívar, Departamento de Física,
\\ AP 89000, Caracas 1080-A, Venezuela}
\email{arestu@usb.ve}
\date{\today}

\begin{abstract}  
We employ  the non-linear realization techniques to relate the
$\mathcal{N}=1$ chiral, and the $\mathcal{N}=2$  vector multiplets to
the Goldstone spin $1/2$ superfield arising from partial supersymmetry
breaking of  $\mathcal{N}=2$ and $\mathcal{N}=3$ respectively. In both
cases, we obtain a family of non-linear transformation laws realizing an
extra supersymmetry. In the ${\mathcal{N}}=2$ case, we find an invariant 
action which  is the low energy limit of the supersymmetric Born-Infeld theory 
expected to describe a D3-brane in six dimensions.   
\end{abstract}

\maketitle

\section{Introduction}  
During the second superstring revolution, we learned that the
$Dp$-branes are the natural Ramond-Ramond  charged objects
\cite{Polchinski:1995mt}. This fact has revealed the importance of the
Born-Infeld action and its possible supersymmetric extensions in diverse
dimensions within the framework of string theory. Besides, the
appearance of Yang-Mills in the effective $Dp$-brane stacks theories
also set as a goal the consistent realization of Born-Infeld dynamics in
the non-Abelian case. In this sense, the last years bore witness to the
second Born-Infeld revolution. It has been conjectured
\cite{Ivanov:2001gd, Bellucci:2001hd} that supersymmetric Born-Infeld
theories emerge naturally form partial supersymmetry breaking (PSSB), as
it has been proposed and successfully tested in several works for
particular cases spanning different dimensions and number of
supersymmetries \cite{Bagger:1997pi, Bagger:1997px, Bagger:1997wp,
Ketov:1998ku, Ivanov:2000nk, Bellucci:2000ft, Kuzenko:2000uh,
Bellucci:2001hd, Ivanov:2001gd, Ivanov:2002zz}. The usual procedure
consists in imposing irreducibility conditions on the Goldstone fields
arising from the PSSB to select the particular representation of the
unbroken residual symmetry in which they lie.  Former studies assigned
this Goldstone superfield to the ${\mathcal{N}}=1$ chiral
\cite{Bagger:1994vj} and Maxwell \cite{Bagger:1997wp} multiplets. In the
${\mathcal{N}}=2$ case, the vector supermultiplet has also been
interpreted as the Goldstone multiplet coming from two broken
supersymmetries \cite{Bellucci:2000ft, Ivanov:2002zz}. Particularly in
\cite{Bellucci:2001hd}, the authors make use of the Goldstone bosonic
$\mathcal{N}=2$ and $\mathcal{N}=4$ superfields associated with the
breaking of $\mathcal{N}=4$ and $\mathcal{N}=8$ central charge
generators respectively. The PSSB mechanism has been also studied within
the framework of M-theory and $d=11$ supergravity using the
superembedding techniques \cite{Pasti:2000zs}.

The common mathematical framework used to describe PSSB is the
non-linear realization method originally developed by Callan, Coleman,
Wess, and Zumino \cite{Coleman:1969sm,Callan:1969sn} for semi-simple
internal symmetries, and extended to supersymmetry by Akulov and Volkov
\cite{Volkov:1972jx}.  In this work we use non-linear realizations, in a
complete off-shell fashion, to derive a general family of non-linear
transformation laws for the chiral Goldstone superfield coming from PSSB
of $\mathcal{N}=2$ supersymmetry without central charges down to
$\mathcal{N}=1$. As special cases, we find the
action obtained by  Bagger and Galperin \cite{Bagger:1994vj}, and the one obtained by Ro\v{c}ek 
and Tseytlin \cite{Rocek:1997hi} up to fourth order in the chiral 
superfield which is the translational invariant action of the $\mathcal{N}=1$ 
3-brane proposed by Hughes and Polchinski in \cite{Hughes:1986dn}.

Later on we deal with the PSSB in the $\mathcal{N}=3\longrightarrow
\mathcal{N}=2$ case, where we obtain a ${\mathcal{N}}=2$ action for the
vector multiplet which is invariant under the extra hidden
supersymmetry. We impose the proper covariant irreducibility conditions
over the spin $1/2$ Goldstone superfield to relate it to the
$\mathcal{N}=2$ chiral superfield containing the vector multiplet.  In
this case we obtain also a family of transformation laws as well as
their invariant actions. In a particular case, the Lagrangian has
exactly the terms that come from the Born-Infeld low energy limit up to
$\Oh(F^6)$. This action has been interpreted as the world-volume
dynamics of a D3-brane propagating in six dimensions
\cite{Tseytlin:1999dj}. The properties required for such model have been
studied in \cite{Kuzenko:2000uh}.  It is important to remark that our
derivation is in both cases  off-shell.

%
%New section
%
\section{Non-linear Realizations in Superspace 
\label{section:NLRS}}
In this section, we briefly review the standard non-linear realizations
formalism for the ${\mathcal{N}}=2\longrightarrow \mathcal{N}=1$ case.
Let $G$ be the super-Poincaré ${\mathcal{N}}=2$ group which we would
like to breakdown and $H\subset G$ the unbroken invariant subgroup.
Consider the splitting of the generators of $G$ into three classes: The
generators of superspace translations
$\Gamma_{\underline{A}}=(P_m,Q_\alpha,\bar{Q}^{\dot\alpha})$, where
$Q_\alpha$ are the residual supersymmetry charges, the broken symmetry
generators $\Gamma_r=(S_\alpha,\bar{S}^{\dot\alpha})$, and the preserved
symmetry generators $\Gamma_i$ belonging to the Lie algebra of $H$ which
corresponds in our case to the Lorenz subgroup $SO(3,1)$.

Starting from the ${\cal{N}}=2$ algebra without central charges
\begin{equation}
\begin{split}
\{Q_\alpha,\bar{Q}_{\dot{\beta}}\}&=2\sigma^m_{\alpha\dot{\beta}}P_m,\\
\{S_\alpha,\bar{S}_{\dot{\beta}}\}&=2\sigma^m_{\alpha\dot{\beta}}P_m,\\
\{Q_\alpha,\bar{S}_{\dot{\beta}}\}&=0,\\ \{Q_\alpha,S_\beta\}&=0,
\end{split}
\end{equation}
the parameterization of the coset space $G/H$ is taken to be
\begin{equation}
\Omega=\exp(\iim
X^{\underline{A}}\Gamma_{\underline{A}})\exp(\iim\psi^\alpha
S_\alpha+\iim\bar{\psi}_{\dot{\alpha}}\bar{S}^{\dot{\alpha}}).
\end{equation} 
Here $\psi^\alpha=\psi^\alpha(x,\theta,\bar\theta)$ is the Goldstone
\emph{superfield} related to the breaking of the $S_\alpha$ generator.

The action of $g=\exp(\iim\eta S+\iim\bar{\eta}\bar{S})$, where
$\eta^\alpha$ is a constant spinor parameter, over a general element of
the coset space can be understood as a transformation law for the
Goldstone fields and the coordinates
\begin{equation}
\begin{split}
(x^{\prime\,m},\theta',\bar{\theta}')
&=(x^m+\iim(\eta\sigma^m\bar{\psi}-\psi\sigma^m\bar{\eta}),\theta,\bar{\theta}),\\
\psi'&=\psi+\eta,\\
\bar\psi'&=\bar\psi+\bar\eta,
\end{split}
\end{equation}
from which we can derive the non-linear transformation law for the
Goldstone field
\begin{equation}\label{GoldNLlaw}
  \delta\psi_\alpha=\eta_\alpha
  -\iim(\eta\sigma^m\bar{\psi}-\psi\sigma^m\bar{\eta})\partial_m\psi.
\end{equation}
We are now able to deduce the covariant ${\mathcal{N}}=1$ 1-forms
\begin{equation}\label{Cov1forms}
\begin{split}
\omega^m(P)&=dx^m+\iim(d\theta\sigma^m\bar{\theta}-\theta\sigma^md\bar{\theta})
+\iim(d\psi\sigma^m\bar{\psi}-\psi\sigma^md\bar{\psi}),\\
\omega^\alpha(Q)&=d\theta^\alpha,\\
\bar{\omega}_{\dot{\alpha}}(\bar{Q})&=d\bar{\theta}_{\dot{\alpha}},\\
\omega^\alpha(S)&=d\psi^\alpha,\\
\bar{\omega}_{\dot{\alpha}}(\bar{S})&=d\bar{\psi}_{\dot{\alpha}},
\end{split}
\end{equation}
that assemble into the invariant Maurer-Cartan 1-form 
\begin{equation}
\begin{split}
\Omega^{-1}d\Omega&=\iim[\omega^m(P)P_m+\omega^\alpha(Q)Q_\alpha+
\bar{\omega}_{\dot{\alpha}}(\bar{Q})\bar{Q}^{\dot{\alpha}}\\ &+
\omega^\alpha(S)S_\alpha+
\bar{\omega}_{\dot{\alpha}}(\bar{S})\bar{S}^{\dot{\alpha}}].
\end{split}
\end{equation}
The expansion of the first three 1-forms in equation (\ref{Cov1forms})
$\omega^{\underline{A}}=dX^{\underline{M}}E_{\underline{M}}^{\underline{A}}$
in terms of
$dX^{\underline{A}}=(dx^m,d\theta^\alpha,d\bar{\theta}_{\dot{\alpha}})$,
lead us to the supervierbein 
\begin{equation}
\begin{split}
E_n^m&=\delta_n^m+ \iim(\partial_n\psi\sigma^m\bar{\psi}-\psi\sigma^m\partial_n\bar{\psi}),\\
E_\beta^m&=\iim(\sigma^m\bar{\theta})_\beta +\iim(\partial_\beta\psi\sigma^m\bar{\psi}+
\psi\sigma^m\partial_\beta\bar{\psi}),\\
E^{\dot{\beta}m}&=\iim(\bar{\sigma}^m\theta)^{\dot{\beta}} +\iim(\bar{\partial}^{\dot{\beta}}
\psi\sigma^m\bar{\psi}+ \psi\sigma^m\bar{\partial}^{\dot{\beta}}\bar{\psi}),\\
E_\beta^\alpha&=\delta_\beta^\alpha,\\
E^{\dot{\beta}}_{\dot{\alpha}}&=\delta^{\dot{\beta}}_{\dot{\alpha}}.
\end{split}
\end{equation}
In the same fashion one must expand the remaining 1-forms
$\omega^\alpha=\omega^{\underline{M}}\mathcal{D}_{\underline{M}}\psi^\alpha$,
to find the covariant derivatives of the Goldstone superfield
$\mathcal{D}_{\underline{M}}\psi^\alpha=
E^{-1\,\underline{N}}_{\underline{M}}\partial_{\underline{N}}\psi^\alpha$.
This calculation involves the inversion of the supervierbein matrix
\begin{equation}
E_{\underline{M}}^{-1\,\underline{N}}=
  \begin{pmatrix}
    \omega_m^{-1\,n} & -E_\alpha^p\omega^{-1\,n}_p & -E^{\dot{\alpha}p}\omega^{-1\,n}_p \\
    0 & \delta_\alpha^\beta & 0 \\
    0 & 0 & \delta^{\dot{\alpha}}_{\dot{\beta}}
  \end{pmatrix},
\end{equation}
which is given in terms of the direct supervierbein and the inverse of
$\omega_n^m=\delta_n^m+
\iim(\partial_n\psi\sigma^m\bar{\psi}-\psi\sigma^m\partial_n\bar{\psi})$.
With these elements we can express the covariant derivatives in a
seemingly explicit form:
\begin{equation}
\begin{split}
\mathcal{D}_m&=\omega_m^{-1\,n}\partial_n,\\
\mathcal{D}_\alpha&=D_\alpha-\iim(D_\alpha\psi\sigma^m\bar{\psi} +\psi\sigma^mD_\alpha\bar{\psi})
\mathcal{D}_m,\\
\mathcal{\bar{D}}_{\dot{\alpha}}&=\bar{D}_{\dot{\alpha}}-\iim (\bar{D}_{\dot{\alpha}}
\psi\sigma^m\bar{\psi} +\psi\sigma^m\bar{D}_{\dot{\alpha}}\bar{\psi})\mathcal{D}_m,
\end{split}
\end{equation}
where $D_\alpha$ and $\bar{D}_{\dot{\alpha}}$ are the usual superspace
derivatives.  (See Appendix \ref{A} for conventions).  By defining
irreducibility constraints over chiral fields with the aid of covariant
derivatives we will be able to treat the ghost states problem of PSSB
requiring the Goldstone multiplet to be an ${\cal{N}}=1$ irreducible
\textit{and} ${\cal{N}}=2$ covariant representation. 

%
%New section
%
\section{The family of non-linear Transformation Laws \\and 
the $\mathcal{N}=1$ Invariant Actions \label{section:family}}
Generally, the chiral multiplet can be associated with the Goldstone
spin-1/2 field imposing constraints on the components of a ${\cal{N}}=1$
SUSY irrep \cite{Rittenberg:1981cp,Sokatchev:1981my}
\begin{equation}
  \bar{D}\bar{D}\psi_\alpha,\qquad \bar{D}\bar{D}D_\beta\psi_\alpha,
  \qquad \bar{D}\bar{D}DD\psi_\alpha.
\end{equation}
To specifically cancel out the irrelevant pieces of the superfield
keeping only the chiral part one must choose \cite{Bagger:1997px}
\begin{equation}
\begin{split}
\bar{D}\bar{D}\psi_\alpha&=0,\\
D_\alpha\psi_\beta + D_\beta\psi_\alpha&=0,
\end{split}\label{IrrPlanoN1}
\end{equation}
as to select only the antisymmetric term
$\bar{D}\bar{D}D_{[\beta}\psi_{\alpha]}$ leading to the chiral
multiplet. This constraint is readily solved by taking
$\psi_\alpha=\lambda D_\alpha\Phi$, where $\Phi$ is a chiral field which
non-linearly realize a second supersymmetry if its chirality condition
is expressed in terms of covariant derivatives.  The parameter $\lambda$
of geometrical dimension $[\lambda]=-2$ will be related to the scale of
supersymmetry breaking. The curve chirality condition can be written as
\begin{equation}\label{ccurva}
\mathcal{\bar{D}}_{\dot{\alpha}}\Phi=0,
\end{equation}
This solves the proper constraints (\ref{IrrPlanoN1}) in the flat limit
$\mathcal{D}\longrightarrow D$.  The solution of this generalized
constraint is defined up to a chiral field $\bar{D}²\mathcal{X}$.
\begin{equation}\label{soluciongeneral}
\begin{split}
\Phi&=\varphi-\iim\lambda^2(D\varphi\sigma^m\bar{D}\bar{\varphi})
\partial_m\varphi
+4\lambda^2\bar{\varphi}(\partial\varphi)^2\\
&+\lambda^2\bar{D}^2\mathcal{X}
+\Oh(\lambda^4).
\end{split}
\end{equation}
$\mathcal{X}$ has geometrical dimension $[\mathcal{X}]=4$ and will be
constructed out of $\varphi$ to preserve $\mathcal{N}=1$ supersymmetry.
Non-linear transformation laws for the fields can then be determined up
to a superfield $Z_\xi$ containing both $\varphi$ and the parameter of
the transformation $\xi$.
\begin{subequations}
\begin{align}
\delta\Phi&=\theta\xi
-\iim\lambda²(\xi\sigma^m\bar{D}\bar\varphi
-D\varphi\sigma^m\bar\xi)\partial_m\varphi+\Oh(\lambda^4),\\
\nonumber\delta\varphi&=\theta\xi
+2\iim\lambda²(D\varphi\sigma^m\bar\xi)\partial_m\varphi
-4\lambda²(\partial\varphi)²(\bar\theta\bar\xi)\\ 
\label{Chiral1Transf}&-\lambda²\bar{D}²\delta \mathcal{X}+\lambda²D²Z_\xi+\Oh(\lambda^4).
\end{align}
\end{subequations}
Considering the liberty of choice of $\mathcal{X}$ and $Z_\xi$ we
actually have a family of non-linear transformation laws, which together
with the fields, have to fulfill the irreducibility constraints at least
in the linear limit. In addition, the fields $\mathcal{X}$ and $Z_\xi$
must also be constrained so that the algebra of the non-linear
transformations correspond to that of an extra supersymmetry. This
amounts to the following equations:
\begin{subequations}
\begin{eqnarray}
\label{N1Constr1}\bar{D}²[\delta_\xi,\delta_\eta]\mathcal{X}&=&D²\delta_{[\xi}Z_{\eta]},\\
	\label{N1Constr2}\bar{D}_{\dot\alpha}D²Z_\xi&=&0.
\end{eqnarray}
\end{subequations}
These conditions  set $Z_\xi=0$ (see Appendix \ref{XZN=1}) but allow
$\mathcal{X}$ to be a very general field. Some admissible solutions are
\begin{equation}\label{xgeneral}
\begin{split}
\mathcal{X}&=a_1\varphi(D\varphi)²+a_2\bar\varphi(D\varphi)²
+a_3\varphi\bar\varphi\bar{D}²\bar\varphi\\&+a_4\bar\varphi(\bar{D}\bar\varphi)²
+a_5\varphi²D²\varphi+a_6\varphi\bar\varphi D²\varphi
\\&+a_7\bar\varphi²D²\varphi+a_8\bar\varphi²\bar{D}²\bar\varphi,
\end{split}
\end{equation}
the superalgebra corresponds to
\begin{equation}
	[\delta_\xi,\delta_\eta]\varphi=
	2\iim\lambda²(\xi\sigma^m\bar\eta-\eta\sigma^m\bar\xi)\partial_m\varphi
	+\Oh(\lambda^4).
\end{equation}
The parameters $a_i$ span the set of superfields defining a family of
non-linear transformation laws.  Now we are ready to build a
$\mathcal{N}=1$ action with a non-linearly realized extra supersymmetry.
We start with the usual $\mathcal{N}=1$ chiral action
\begin{equation}
	S_1=\int d^4x\,d²\theta\,d²\bar\theta\,\varphi\bar\varphi.
	\label{N1Chiral}
\end{equation}
Its variation with respect to the transformation law (\ref{Chiral1Transf}) is then
\begin{multline}
	\delta S_1=\int d^4x\,d²\theta\,d²\bar\theta\,
	\Bigl[(\theta\xi)\bar\varphi
	+2\iim\lambda²(D\varphi\sigma^m\bar\xi)\partial\varphi\bar\varphi\\
	-4\lambda²(\partial\varphi)²(\bar\theta\bar\xi)\bar\varphi
	-\lambda^2\bar{D}²\delta \mathcal{X}\bar\varphi\Bigr] + \text{c.c.} + \Oh(\lambda^4).
\end{multline}
Due to the fact that $(\theta\xi)\bar\varphi$,
$\bar{D}²\mathcal{X}(\bar\theta\bar\xi)$ and their  complex conjugates
do not contribute to the integral above, we can express the variation of
$S_1$ as the first order variation of terms of second order in $\lambda$
\begin{multline} 
	\delta S_1=\lambda²\delta\int
	d^4x\,d²\theta\,d²\bar\theta\, 
	\Biggl[\frac14(D\varphi)²(\bar{D}\bar\varphi)²\\
	-2(\partial\varphi)²\bar\varphi² 
	-2(\partial\bar\varphi)²\varphi² 
	+\bar{D}²\mathcal{X}\bar\varphi
	+D²\bar{\mathcal{X}}\varphi\Biggr]+\Oh(\lambda^4).  
	\label{varN1Chiral} 
\end{multline} 
Then it is a simple task to propose a set of  $\mathcal{N}=1$ chiral
gauge actions with an extra non-linear supersymmetry
\begin{multline} 
	\hat{S}_1=\int d^4x\,d²\theta\,d²\bar\theta\, 
	\Bigg[\varphi\bar\varphi +2\lambda²(\partial\varphi)²\bar\varphi²
	+2\lambda²(\partial\bar\varphi)²\varphi²\\ 
	-\frac14\lambda²(D\varphi)²(\bar{D}\bar\varphi)²
	+\lambda²\bar{D}²\mathcal{X}\bar\varphi 
	+\lambda²D²\bar{\mathcal{X}}\varphi \Biggr]  
	+ \Oh(\lambda^4). 
	\label{NLN=1+1} 
\end{multline} 
Choosing $\mathcal{X}=0$ we obtain the action of Bagger and Galperin
\cite{Bagger:1994vj}, coming from the breaking of $\mathcal{N}=2$
supersymmetry with central charges.  Note that our conventions are
different from the cited paper  (see Appendix \ref{A}). 
We can further impose phase invariance of the action
\begin{equation}
	a_1=a_3=a_4=a_5=a_7=a_8=0.
\end{equation}
The $N=1$ chiral action (\ref{NLN=1+1}) with this constraints will then be equivalent to
\begin{widetext}
	\begin{multline}
	\hat{S}_1=\int d^4x\,d²\theta\,d²\bar\theta\,
	\Bigg[\varphi\bar\varphi%\\
	+2(1+4a_2+8a_6)\lambda²\left[(\partial\varphi)²\bar\varphi²
	+(\partial\bar\varphi)²\varphi²\right]\\
	-\left(\frac14+2a_2-2a_6\right)\lambda²(D\varphi)²(\bar{D}\bar\varphi)²%\\
	+\frac14a_6\lambda²\bar{D}²\bar\varphi²D²\varphi²
	\Biggr]+\Oh(\lambda^4).
\end{multline}
\end{widetext}
Choosing $a_2=-\frac{1}{4}$ and $a_6=0$ we find the action in \cite{Rocek:1997hi} up to $\Oh(\varphi^4)$ which is the translational invariant action of the $\mathcal{N}=1$ 3-brane proposed in \cite{Hughes:1986dn}. It is important to comment that all these actions are equivalent through field redefinitions. Nevertheless, we have used the generalized family of transformation laws (\ref{Chiral1Transf}) to obtain the actions in \cite{Bagger:1994vj} and \cite{Rocek:1997hi}. Our analysis suggests  that the nature of the transformation laws (\ref{Chiral1Transf}) is the origin of the equivalence between these dynamics.
%New section
%
\section{Partial ${\cal{N}}=3$ Supersymmetry Breaking
\label{section:PSSB}} 
In this section, we construct a ${\mathcal{N}}=2$ action  for the vector
multiplet which is invariant under a third non-linearly realized
supersymmetry. This action contains  non-linear terms in the spin one
gauge fields which are interpreted as the low energy limit of the
supersymmetric Born-Infeld theory.  We repeat the procedure of section
\S \ref{section:NLRS} but enlarging the superspace by one Grassmann
variable $(X^m, \theta_A)$ with $\theta_A=(\theta, \tilde\theta)$. This
notation will be understood for every Grassmann variable or operator
i.e. $D^A_\alpha=(D_\alpha, \tilde{D}_\alpha)$. Indexes $A,B$ are
$SU(2)$ and therefore raised and lowered with $\epsilon_{AB}$.
We will represent the volume element in the Grassmann variables by
\begin{equation} 
	\begin{aligned} 
		d\Theta&=d²\theta d²\tilde{\theta},
		&d\bar{\Theta}&=d²\bar{\theta}d²\bar{\tilde{\theta}},\\
		d^4\theta&=d²\theta d²\bar\theta, 
		&d^4\tilde\theta&=d²\tilde\theta d²\bar{\tilde\theta}.
	\end{aligned} 
\end{equation} 
In this case the total group $G$ is the $\mathcal{N}=3$ supersymmetry
group  and $H$ is still $SO(3,1)$ The ${\cal{N}}=3$ algebra without
central charges is 
\begin{equation} 
	\begin{split} 
		\{Q^A_\alpha,\bar{Q}_{\dot{\beta}B}\}
		&=2\delta^A_B\sigma^m_{\alpha\dot{\beta}}P_m,\\
		\{S_\alpha,\bar{S}_{\dot{\beta}}\}&=2\sigma^m_{\alpha\dot{\beta}}P_m,\\
		\{Q^A_\alpha,\bar{S}_{\dot{\beta}}\}&=0,\\
		\{Q^A_\alpha,S_\beta\}&=0, 
	\end{split} 
\end{equation} 
the parameterization of the coset space $G/H$ has the same form as the
former case 
\begin{equation} 
	\Omega=\exp(\iim X^r\Gamma_r)\exp(\iim\psi^\alpha S_\alpha
	+\iim\bar{\psi}_{\dot{\alpha}}\bar{S}^{\dot{\alpha}}).
\end{equation} 
Now $\psi^\alpha=\psi^\alpha(x,\theta_A,\bar{\theta}^A)$ is the
Goldstone \emph{superfield} related to the breaking of the $S_\alpha$
generator.  Though the transformation law for the Goldstone superfield
is analogous to (\ref{GoldNLlaw}), the supervierbein is in this case 
\begin{equation} 
	\begin{split}
		E_n^m&=\delta_n^m+
		\iim(\partial_n\psi\sigma^m\bar{\psi}-\psi\sigma^m\partial_n\bar{\psi}),\\
		E_\beta^{m A}&=\iim(\sigma^m\bar{\theta}^A)_\beta
		+\iim(\partial^A_\beta\psi\sigma^m\bar{\psi}
		+\psi\sigma^m\partial^A_\beta\bar{\psi}),\\ E^{\dot{\beta}m}_
		A&=\iim(\bar{\sigma}^m\theta_A)^{\dot{\beta}}
		+\iim(\bar{\partial}^{\dot{\beta}}_A\psi\sigma^m\bar{\psi}+
		\psi\sigma^m\bar{\partial}^{\dot{\beta}}_A\bar{\psi}),\\
		E_\beta^{\alpha}&=\delta_\beta^\alpha,\quad
		\tilde{E}_\beta^{\alpha}=\delta_\beta^\alpha,\\
		E^{\dot{\beta}}_{\dot{\alpha}}&=\delta^{\dot{\beta}}_{\dot{\alpha}},
		\quad
		\tilde{E}^{\dot{\beta}}_{\dot{\alpha}}=\delta^{\dot{\beta}}_{\dot{\alpha}}.
	\end{split} 
\end{equation} 
The covariant non-linear derivatives of the Goldstone superfield
$\mathcal{D}^A_{\underline{M}}\psi^\alpha=
(E^{A\,\underline{N}}_{\underline{M}})^{-1}\partial_{\underline{N}}\psi^\alpha$
are 
\begin{equation} 
	\begin{split}
		\mathcal{D}_m&=\omega_m^{-1\,n}\partial_n,\\
		\mathcal{D}_\alpha^A&=D^A_\alpha-\iim(D^A_\alpha\psi\sigma^m\bar{\psi}
		+\psi\sigma^mD^A_\alpha\bar{\psi}) \mathcal{D}_m,\\
		\mathcal{\bar{D}}_{\dot{\alpha}A}&=\bar{D}_{\dot{\alpha}A}-\iim
		(\bar{D}_{\dot{\alpha}A} \psi\sigma^m\bar{\psi}
		+\psi\sigma^m\bar{D}_{\dot{\alpha}A}\bar{\psi})\mathcal{D}_m.
	\end{split} 
\end{equation}
Starting with the ${\mathcal{N}}=2$ superfield
${\mathcal{W}}={\mathcal{W}}(x,\theta^A,\bar{\theta}^B)$, it is possible
to build an irreducible representation of the ${\cal{N}}=2$
supersymmetry in terms of ${\cal{N}}=1$ superfields by imposing the
following chirality and reality conditions \cite{Grimm:1978xp},
\begin{align}\label{chirality-reality}
	\bar{D}^A_{\dot\alpha}{\mathcal{W}}=0, \quad
	D^{\alpha\,A}{D}^B_{\alpha}{\mathcal{W}}
	+\bar{D}_{\dot\alpha}^A\bar{D}^{\dot\alpha\,B}\bar{\mathcal{W}}=0.
\end{align} 
Performing the appropriate translation
$y^m=x^m-\iim\theta_A\sigma^m\bar{\theta}^A$, and imposing the
aforementioned chirality and reality conditions, we obtain an expansion
of ${\mathcal{W}}$ in terms of ${\cal{N}}=1$ superfields
\begin{equation} 
	{\mathcal{W}}=\phi(y,\tilde\theta)
	+\sqrt2\theta^{\alpha} W_\alpha(y,\tilde\theta) 
	+\theta\theta G(y,\tilde\theta), 
\end{equation} 
where 
\begin{equation}
	G(y,\tilde{\theta}) =\int d²\bar{\tilde{\theta}}\,\bar{\phi}
	(y+\iim\tilde{\theta}\sigma\bar{\tilde{\theta}},\bar{\tilde{\theta}})
	e^{2V(y+\iim\tilde{\theta}\sigma\bar{\tilde{\smash\theta}},
	\bar{\tilde{\smash\theta}})}, 
\end{equation} 
and $V$ is the prepotential of
$W_\alpha=\iim\bar{\widetilde{D}}²\widetilde{D}_\alpha V$.  These
superfields transform in the usual way under the $\tilde{Q},
\bar{\tilde{Q}}$ generators but transform into each other under
$Q,\bar{Q}$ as components of a chiral multiplet, in other words
according to 
\begin{equation}\label{N2Law}
	\begin{aligned} 
		\delta \phi &=\sqrt2(\xi W),\\ 
		\delta W_\alpha&=
		-\iim\sqrt2(\sigma^m\bar{\xi})_\alpha\partial_m\phi+
		\sqrt2\xi_\alpha G,\\ 
		\delta G&=\iim\sqrt2\partial_mW\sigma^m\bar{\xi}.
	\end{aligned}
\end{equation} 
The usual free ${\mathcal{N}}=2$ super-Maxwell action constructed with
the ${\mathcal{N}}=2$ superfield is 
\begin{equation} 
	S_2=\int d^4x\,d\Theta\,{\mathcal{W}}²
	+\int d^4x\,d\bar{\Theta}\,\bar{{\mathcal{W}}}².\label{N2Free} 
\end{equation} 
Integrating over $\theta$ we obtain the following Lagrangian density
\begin{equation}\label{N2SMaxwell} 
	\frac14\int	d²\tilde{\theta}\,W^2
	+\frac14\int d²\bar{\tilde\theta}\,\bar{W}^2 
	-\int d^4\tilde{\theta}\,\bar{\phi}e^{2V}\phi,
\end{equation} 
which is manifestly ${\mathcal{N}}=1$, gauge ($\delta
V=\iim(\Lambda-\bar{\Lambda})$) invariant and ${\mathcal{N}}=2$ via the
transformation law (\ref{N2Law}). To relate the Goldstone supermultiplet with the $\mathcal{N}=2$
superfield we proceed in analogy to the previous case, identifying
\begin{equation} 
	\psi_\alpha=\lambda\mathcal{D}_\alpha\Phi,
	\label{Powerseries} 
\end{equation} 
and building a set of covariant constraints that reduce to
(\ref{chirality-reality}) in the flat limit 
\begin{subequations} 
	\begin{align}
		\bar{\mathcal{\widetilde{D}}}_{\dalpha}\Phi&=0,\\
		\bar{\mathcal{D}}_{\dalpha}\Phi+
		\frac{1}{4}\lambda^2\bar{\mathcal{D}}_{\dalpha}
		\bar\Phi\bar{\mathcal{D}}^2(D\Phi)^2&=0.  
	\end{align}
\end{subequations} 
This covariant chirality conditions are solved by
\begin{equation} 
	\Phi = \mathcal{W}	
	-\iim\lambda²(D\mathcal{W}\sigma^m\bar{D}\bar{\mathcal{W}})
	\partial_m\mathcal{W}
	+\lambda²X+\Oh(\lambda^4).
	\label{PowerPhi} 
\end{equation}
Where $X$ is some $\mathcal{N}=2$ chiral superfield
$\bar{D}_{\dot\alpha}X=\bar{\tilde{D}}_{\dot\alpha}X=0$.  Considering
that $\lambda$ has dimensions of $L^{-2}$, $X$ cannot be trivially
$\mathcal{W}$, instead $X$ is of third order in $\mathcal{W}$.  As it
has been pointed out in other cases \cite{Bagger:1997wp}, the existence
of dimensionless invariants
$\bar{\mathcal{D}}^A_{\dot\alpha}\mathcal{D}_\beta\Phi$ and
$\mathcal{D}^A_\alpha\mathcal{D}_\beta\Phi$ has a direct impact on the
uniqueness of the constraints. In principle we could add any power of
the dimensionless invariants without spoiling the flat limit
$\lambda\rightarrow0$.  Keeping this in mind, we see that the remaining
reality conditions, are generalized to
 \begin{equation} 
	{\mathcal{D}}^{\alpha\,A}{\mathcal{D}}^B_{\alpha}{\Phi}
+{\mathcal{\bar{D}}}_{\dot\alpha}^A{\mathcal{\bar{D}}}^{\dot\alpha\,B}\bar{\Phi}
-\lambda^2 ({\mathcal{D}}^{\alpha\,A}{\mathcal{D}}^B_{\alpha}X
+{\mathcal{\bar{D}}}_{\dot\alpha}^A{\mathcal{\bar{D}}}^{\dot\alpha\,B}\bar{X})+\lambda^2 f^{AB}(\Phi,\bar{\Phi})=0	
	\label{Gen-reality} 
\end{equation}
where $f^{AB}(\Phi,\bar{\Phi})$ is a function determined by an iterative procedure such that (\ref{PowerPhi}) solves indeed the former conditions. We notice that $X$ is so far only restricted by the chirality conditions but not by the reality ones. The restrictions (\ref{Gen-reality}) reduce in the $\lambda\rightarrow\, 0$ limit to the usual reality conditions of the $\mathcal{N}=2$  field $\mathcal{W}$ defined by
(\ref{chirality-reality}). The non-linear transformation law that
realize a third supersymmetry on $\Phi$ can be derived from
(\ref{GoldNLlaw}) and (\ref{Powerseries}) 
\begin{equation}\label{transformacion}
	\delta\Phi=\theta\xi 
	-\iim\lambda²(\xi\sigma^m\bar{D}\bar\Phi
	-D\Phi\sigma^m\bar\xi)\partial_m\Phi+\Oh(\lambda^4).
\end{equation} 
This transformation law fulfills the constraints (\ref{Gen-reality})
above and, as in the former case, closes in the supersymmetry algebra 
\begin{equation} 
	[\delta_\xi,\delta_\eta]\Phi=
	2\iim\lambda²(\xi\sigma^m\bar\eta-\eta\sigma^m\bar\xi)\partial_m\Phi
	+\Oh(\lambda^4).  
\end{equation} 
Choosing $X=0$, an invariant action for $\Phi$ is
\begin{equation}\label{laaccion} 
	S=\int d^4x\, d\Theta\,\left[ \Ber\Phi^2
	-2\iim\lambda² (\partial_mD\Phi\sigma^m\bar{D}\bar\Phi)\, \Phi^2\right]
	+\frac{a\lambda^2}{2}\int{d^4x\,d\Theta\,d\bar{\Theta}
	\Phi^2\bar{\Phi}^2}+\text{c.c.}
\end{equation} 
Here the Berezinian is included up to order $\lambda²$ 
 \begin{equation}
\Ber=1+\iim\partial_m\psi\sigma^m\bar{\psi}-\iim \psi\sigma^m\partial_m\bar{\psi}.
\end{equation}
Where $a$ is a constant factor. The reason to introduce the additional term in the action beyond the one involving the Berezinian is the presence of the term $\theta\xi$ in the transformation law of $\Phi$ which is not a superfield under the linear supersymmetry transformations. For $a=-\frac{1}{4}$, the action (\ref{laaccion}) is invariant under the linearized $\mathcal{N}=1$ and $\mathcal{N}=2$ supersymmetries and under the transformation law (\ref{transformacion}), which in terms of the $\mathcal{N}=2$ chiral superfield reads  
\begin{widetext} 
\begin{equation}\label{AccionInvN2}
	\hat{S}_2=\int d^4x\,d\Theta\,{\mathcal{W}}² +\int
	d^4x\,d\bar\Theta\,\bar{\mathcal{W}}^2 -\frac{1}{4}\lambda^2\int
	d^4x\,d\Theta\,d\bar\Theta\,
	{\mathcal{W}}^2\bar{\mathcal{W}}^2+\Oh(\lambda^4).
	\end{equation} 
\end{widetext} 
The action (\ref{AccionInvN2}) is self dual in our
approximation. Moreover, (\ref{AccionInvN2}) is  the  $\mathcal{N}=2$ supersymmetric low energy Born-Infeld action, that coincides up to fourth order in the superfields with that proposed by Ketov \cite{Ketov:1998sx,Ketov:2000zw}, as the
$\mathcal{N}=2$ supersymmetric extension of the 4-dimensional
Born-Infeld action, that found by Bellucci, Ivanov and Krivonos
\cite{Bellucci:2001hd} by PSSB of
$\mathcal{N}=4\longrightarrow\mathcal{N}=2$ with central charges, and
that found by Kuzenko and Theisen \cite{Kuzenko:2000uh} up to order
$\Oh(F^{8})$ by requesting self-duality. Hence this action fulfills the
requirements for describing the dynamics of a single D3-brane in six
dimensions. The whole analysis may be now performed for $X\neq\,0$. 
A general choice of the chiral field $X$ in
terms of $\mathcal{W}$ has the form 
\begin{equation}
	X=b\bar{D}²\bar{\tilde{D}}²\bar{\mathcal{W}}^3, 
\end{equation} 
where $b$ is a dimensionless constant. Including $X\neq\,0$ we obtain the
following action
\begin{widetext} 
\begin{equation}
	\hat{S}_2=\int d^4x\,d\Theta\,{\mathcal{W}}²
	+\lambda^2\int d^4x\,d\Theta\,d\bar\Theta\,
	[a{\mathcal{W}}^2\bar{\mathcal{W}}^2 
	+2b\mathcal{W}\bar{\mathcal{W}}^3]+ \text{c.c.}+\Oh(\lambda^4).
	\label{AccionInvN3} 
\end{equation} 
\end{widetext} 
which is invariant under the non-linear transformation law (\ref{transformacion}).
The parameter $b$ can not be absorbed into $\mathcal{W}$ by the redefinition
\begin{equation*}
\widetilde{\mathcal{W}}=\mathcal{W}+\lambda^2 X
\end{equation*}
since, though $X$ is chiral, it is not restricted by any reality condition so that 
(\ref{AccionInvN3}) comprises a plethora of dynamics. This is not completely unexpected since the curve reality constraints on $\Phi$ are $X$ dependent. 
%
%Conclusions
%
\section{Concluding Remarks} 
In the first part of this work we found a family of non-linear transformation 
laws realizing an extra supersymmetry on the chiral $\mathcal{N}=1$ superfield,
up to second order in
the scale of the supersymmetry breaking, obtained from the non-linear
realizations method.  We imposed some restrictions over the most general
set of variations, asking for the subset that could be considered as
second supersymmetries non-linearly realized, and choosing the ones that
fulfill the $\mathcal{N}=2$ algebra.  Moreover,  we were able to find
the action constructed in \cite{Bagger:1994vj} as a particular case. 
In another case, we find the action in \cite{Rocek:1997hi} up to 
$\Oh(\varphi^4)$ which is the translational invariant action of the 
$\mathcal{N}=1$ 3-brane proposed in \cite{Hughes:1986dn}. Instead of making cumbersome 
field redefinitions, we move through the set of Lagrangians by selecting the values of the parameters $a_i$. In the second part, we constructed a family of $\mathcal{N}=2$  actions
invariant under a third broken hidden supersymmetry, considering the
chiral $\mathcal{N}=2$ superfield as the Goldstone field coming from the
partial supersymmetry breaking of a $\mathcal{N}=3$ theory. This actions
are self dual up to $\Oh(\lambda^2)$.  In a specific case,  the
Lagrangian is the $\mathcal{N}=2$ supersymmetric Born-Infeld up to
$\Oh(F^4)$ that describes the world volume dynamics of a single D3-brane
propagating in six dimensions.  It is important to stand out that this
result and those in the literature \cite{Bagger:1997pi, Bagger:1997px,
Bagger:1997wp, Bellucci:2000ft, Kuzenko:2000uh, Bellucci:2001hd,
Ivanov:2001gd, Ivanov:2002zz}, provide examples of a supersymmetry
Born-Infeld theory that arises from the partial supersymmetry breaking.
It seems to be that the nature of this SUSY non-linear electromagnetic
dynamics comes from the partially breaking of higher supersymmetries. 

Though the non-linear realization formalism was carried out in both
cases completely without central charges, and the geometrical objects
involved are independent of them, the resulting algebras reveal the
presence of a hidden central charge.  Owing to the generality of the
off-shell procedure followed, we presume that for the cases here
studied, it is not possible to break the supersymmetry without central
charges.  These  should be associated to the maximal  automorphism group
\cite{Bagger:1994vj}.  

Due to the preservation of $2/3$ of the supersymmetry, some non-BPS
$D$-brane dynamics could arise from the breaking of
$\mathcal{N}=3\longrightarrow2$, as suggested by E. Ivanov
\cite{Ivanov:Personal}. This situation could already be present in our
family of Lagrangians or in another Goldstone multiplet selection.

In the context of string theory it is also important to find the correct
non-Abelian supersymmetric Born-Infeld functional. Some progress in this
direction has been made in \cite{Ketov:2000fv,Bergshoeff:2001dc}. It is
still unknown if there exists a Goldstone multiplet coming from PSSB
that produces a non-Abelian supersymmetric Born-Infeld theory.

%
%Acknowledgments
%
\begin{acknowledgments}
We are grateful to Pío Arias, Evgenyi Ivanov, Olaf Lechtenfeld and
Isbelia Martín for useful discussions, and Dimitri Sorokin for calling
our attention to important references.  This work is partially supported
by the joint Deutscher Akademischer Austausch Dienst-Fundación Gran
Mariscal de Ayacucho scholarship (L. Quevedo) and fellowship (A.De
Castro.), by Instituto Venezolano de Investigaciones científicas (A. De
Castro)  and by Universidad Simón Bolívar, Decanato de Investigaciones
(A. Restuccia).
\end{acknowledgments}

%
%Appendix
%
\appendix

\section{Notation and Conventions
\label{A}}
We use the following signature for the metric and antisymmetric tensors
\begin{equation}
	\eta_{mn}=\diag(-+++),\qquad 
	\varepsilon_{\alpha\beta}=\varepsilon^{\dot\alpha\dot\beta}
	=\begin{pmatrix}\;0&-1\\\;1&\;\;0\end{pmatrix}.
\end{equation}
The Pauli matrices and superspace derivatives are taken in the following
representation
\begin{equation}
	\sigma^m=\left[-\mathbbm1,\;\;
	\begin{pmatrix}0&\;1\\ \;1&\;0\end{pmatrix},\;\;
	\begin{pmatrix}0&-\iim\\ \; \iim&\;\;0\end{pmatrix},\;\;
	\begin{pmatrix}1&\;\;0\\0&-1\end{pmatrix}\right].
\end{equation}
\begin{equation}
D_\alpha=\partial_\alpha-\iim(\sigma^m\bar\theta)_\alpha\partial_m,\quad
\bar{D}_{\dot\alpha}=-\bar\partial_{\dot\alpha}
+\iim(\theta\sigma^m)_{\dot\alpha}\partial_m.
\end{equation}
The anticommutator between derivatives will then be
\begin{equation}
	\{D_\alpha,\bar{D}_{\dot\beta}\}=2\iim\sigma^m_{\alpha\dot\beta}\partial_m.
\end{equation}
In general, we follow the conventions in \cite{Bagger:1990qh} for the
contraction of the spinorial indexes.
\section{A family of $N=2$ non-linear transformation laws
\label{XZN=1}}
In principle, the most general superfield $Z_\xi$ containing $\varphi$ that 
is dimensionally consistent is 
\begin{widetext}
\begin{equation}
\begin{split}
Z_\xi&=b_1\varphi\xi^\alpha D_\alpha\varphi
+b_2\varphi\bar\xi_{\dot\alpha} \bar{D}^{\dot\alpha}\bar\varphi
+b_3(\theta\xi)(D\varphi)²+b_4(\bar\theta\bar\xi)(D\varphi)²
+b_5\bar\xi_{\dot\alpha} \bar{D}^{\dot\alpha}\bar\varphi
\theta^\alpha D_\alpha\varphi\\
&+b_6\xi^\alpha D_\alpha\varphi\bar\theta_{\dot\alpha}
\bar{D}^{\dot\alpha}\bar\varphi
+b_7(\theta\xi)(\bar{D}\bar\varphi)²
+b_8(\bar\theta\bar\xi)(\bar{D}\bar\varphi)²
+b_9(\theta\xi)D²\varphi\varphi
+b_{10}(\theta\xi)D²\varphi\bar\varphi\\
&+b_{11}(\bar\theta\bar\xi)D²\varphi\varphi
+b_{12}(\theta\xi)\bar{D}²\bar\varphi\varphi
+b_{13}(\theta\xi)\bar{D}²\bar\varphi\bar\varphi
+b_{14}(\bar\theta\bar\xi)\bar{D}²\bar\varphi\varphi\\
&+b_{15}(\bar\theta\bar\xi)\bar{D}²\bar\varphi\bar\varphi
+b_{16}(\theta\sigma^m\bar\xi)\partial_m\varphi\varphi
+b_{17}(\theta\sigma^m\bar\xi)\partial_m\varphi\bar\varphi
+b_{18}(\xi\sigma^m\bar\theta)\partial_m\varphi\varphi\\
&+b_{19}(\xi\sigma^m\bar\theta)\partial_m\varphi\bar\varphi
+b_{20}(\theta\sigma^m\bar\xi)\partial_m\bar\varphi\varphi
+b_{21}(\xi\sigma^m\bar\theta)\partial_m\bar\varphi\varphi
\end{split}
\end{equation}
\end{widetext}
where  $b_j$ are constants to be determined.
It is easy to see that $\mathcal{X}$ of equation (\ref{xgeneral}) satisfies $\bar{D}²[\delta_\xi,\delta_\eta]\mathcal{X}=0$ so according to 
(\ref{N1Constr1}) $Z_\xi$ must satisfy
\begin{equation}
	D²\delta_{[\xi}Z_{\xi]}=0,
\end{equation}
this restricts the field to
\begin{equation}
	\begin{split}
		b_2+b_5+b_{20}&=2b_7,\\
		b_{13}&=b_{14},\\
		b_{16}=b_{17}=b_{18}=b_{19}&=0.
	\end{split}
\end{equation}
The other constraint (\ref{N1Constr2}) cancels the remaining terms in $Z_\xi$. 
As $\mathcal{X}$ is not restricted, we have found in (\ref{Chiral1Transf}) a family of non-linear 
transformations realizing an extra supersymmetry over the chiral $N=1$ action.

\bibliography{32PSSB}% Produces the bibliography via BibTeX.

\end{document}